\documentclass[11pt]{article}
\usepackage{color,epsf}
\usepackage{amssymb,slashed,latexsym,amsmath,multirow,color,pifont}
\usepackage{latexsym,bm,amsfonts}
\usepackage{graphics}
\usepackage{wasysym}

\usepackage{marvosym}
\usepackage{cite}

\makeatletter
\@addtoreset{equation}{section}
\makeatother

\usepackage{float}

\usepackage{soul}

\usepackage[font=footnotesize,labelsep=newline,labelfont=sc,justification=centering,position=top]{caption}

\newcommand{\bea}{\setlength\arraycolsep{2pt} \begin{eqnarray}}
\newcommand{\eea}{\end{eqnarray}}
\newcommand{\nn}{\nonumber}
\newcommand{\cmark}{\ding{51}}
\newcommand{\xmark}{\ding{55}}

\usepackage{hyperref}

\newcommand{\ft}[2]{{\textstyle\frac{#1}{#2}}}
\def\rmi{{\rm i}}
\newsavebox{\uuunit}
\sbox{\uuunit}
    {\setlength{\unitlength}{0.825em}
     \begin{picture}(0.6,0.7)
        \thinlines
        \put(0,0){\line(1,0){0.5}}
        \put(0.15,0){\line(0,1){0.7}}
        \put(0.35,0){\line(0,1){0.8}}
       \multiput(0.3,0.8)(-0.04,-0.02){12}{\rule{0.5pt}{0.5pt}}
     \end {picture}}

\def\be{\begin{equation}}
\def\ee{\end{equation}}
\def\ba{\begin{array}}
\def\ea{\end{array}}
\def\bea{\begin{eqnarray}}
\def\eea{\end{eqnarray}}
\def\bd{\begin{displaymath}}
\def\ed{\end{displaymath}}

\def\nn{\nonumber}

\def\a{\alpha}

\def\e{\epsilon}

\def\vf{\varphi}

\def\l{\lambda}
\def\L{\Lambda}
\def\m{\mu}
\def\n{\nu}

\def\s{\sigma}

\def\nn{\nonumber}

\def\cN{\mathcal{N}}

\begin{document}

\begin{titlepage}

\begin{center}

\vskip 1.5cm

{\Large \bf Supersymmetric Solutions of $\cN = (1,1)$ General Massive Supergravity}
\vskip 1cm

{\bf N.S. Deger\,$^1$, Z. Nazari\,$^2$, {\"O}. Sar{\i}o\u{g}lu\,$^3$.}  \\

\vskip 25pt

\end{center}

{\em $^1$ \hskip -.1truecm Department of Mathematics, 
Bogazici University, 34342, Bebek, Istanbul, Turkey \vskip 5pt }

\vskip 5pt

{\em $^2$ \hskip -.1truecm  Department of Physics, Bogazici University, 34342, Bebek, Istanbul, Turkey
\vskip 5pt }

\vskip 5pt

{\em $^3$ \hskip -.1truecm  Department of Physics,
              Middle East Technical University, 06800, Ankara, Turkey 
							
							\vskip 5pt }

\begin{center}
sadik.deger@boun.edu.tr, zainab.nazari@boun.edu.tr, sarioglu@metu.edu.tr 

\vskip 15pt

\vskip 0.5cm

{\bf ABSTRACT}\\
\end{center}
We construct supersymmetric solutions of three dimensional $\cN = (1,1)$ General Massive Supergravity (GMG). 
Solutions with a null Killing vector are in general pp-waves. We identify those that appear at critical points of the model 
some of which do not exist in $\cN = (1,1)$ New Massive Supergravity (NMG). In the timelike case, we find that many solutions are common with NMG 
but there is a new class that is genuine to GMG, two members of which are stationary Lifshitz and timelike squashed AdS spacetimes. We also show that in addition to the fully supersymmetric AdS vacuum, there is a second AdS background with a non-zero vector field that 
preserves 1/4 supersymmetry.

\end{titlepage}

\newpage

\setcounter{page}{1}

\tableofcontents

\section{Introduction}

Trying to understand quantum gravity in three rather than four dimensional spacetime is a technically more manageable problem.
A reason for this is that, a 3-dimensional gravity theory on anti-de Sitter (AdS) space is dual to a 2-dimensional conformal field theory (CFT)
and such CFTs are much better understood compared to higher dimensional ones. With this goal in mind, Topologically Massive Gravity (TMG) 
\cite{Deser:1981wh} has been studied widely in recent years (see e.g. \cite{Li:2008dq}) which is obtained by adding the gravitational Chern-Simons term
to pure Einstein gravity with or without a cosmological constant. This model is unitary and propagates a single massive mode.
Recently, a novel modification of this theory was achieved where a particular four derivative curvature term was added to the TMG action after which it remains unitary and
there are two massive graviton states. Depending on whether the model contains the gravitational Chern-Simons term or not, it is called General Massive Gravity (GMG) or New Massive Gravity (NMG), respectively \cite{Bergshoeff:2009hq, Bergshoeff:2009aq}. 

Since supersymmetry in general improves ultraviolet behaviour, it is natural to consider supersymmetric extensions of these models which was carried out in a series of papers.
The fact that the isometry group of $AdS_3$ can be written as $SO(2,2) \simeq SO(2,1) \times SO(2,1)$ makes it possible to have $\cN=(p,q)$ supergravities in 
3-dimensions \cite{Achucarro:1987vz} with either on-shell or off-shell formulations.  The off-shell $\cN=1$ \cite{Uematsu:1984zy, Uematsu:1986de}
and $\cN = 2$ \cite{Rocek:1985bk , Nishino:1991sr} TMG models were obtained shortly after the appearance of the bosonic version \cite{Deser:1981wh}. After some gap, 
$\cN=1$ and $\cN=2$ supersymmetric extensions of GMG were constructed in \cite{Andringa:2009yc, Bergshoeff:2010mf} and \cite{Alkac:2014hwa}, respectively. This problem 
has also been studied in superspace \cite{Kuzenko:2011xg, Kuzenko:2011rd, Kuzenko:2013uya, Kuzenko:2015jda}.

Identifying supersymmetric vacua of these theories is an important problem and in this paper we will study supersymmetric solutions of $\cN=(1,1)$ GMG \cite{Alkac:2014hwa}. 
In \cite{Deger:2013yla} a general Killing Spinor analysis was performed to classify supersymmetric solutions of $\cN=(1,1)$ TMG and some particular warped AdS 
solutions were found. A big advantage of working with off-shell supergravities is 
that such an analysis remains valid for any extension of the model, only field equations change. Later, supersymmetric solutions of $\cN=(1,1)$ NMG were examined in
\cite{Alkac:2015lma} and it was found that some additional backgrounds are allowed compared to $\cN=(1,1)$ TMG, such as static Lifshitz spacetime. We will show that
for $\cN=(1,1)$ GMG even more configurations appear such as stationary Lifshitz spacetime and a timelike warped AdS with no restriction on the norm of the warping. The 
latter can be turned into a black hole like object when it is squashed after a proper identification of a 
coordinate\cite{Anninos:2008fx}.
Moreover, we find that in addition to a maximally supersymmetric AdS vacuum, there exists a 1/4 supersymmetric AdS vacuum 
too.  As usual, supersymmetric solutions can be grouped as null or timelike with respect to the norm of a Killing vector that is obtained as a Killing spinor bilinear \cite{Tod:1983pm}.
We summarize our findings for the timelike case in Table \ref{table} where we also make comparison with $\cN=(1,1)$ NMG and TMG models.   In all these papers the same ansatz for the auxiliary fields is assumed.

The organization of our paper is as follows. In the next section we give a brief introduction of $\cN=(1,1)$ GMG model \cite{Alkac:2014hwa}. The two subsequent sections 
constitute our main results where we apply Killing spinor analysis of \cite{Deger:2013yla} to our model. In section \ref{Null}, we analyze a null Killing vector and show that only
pp-waves are allowed. We also explicitly give new solutions that appear at critical points of the theory. In section \ref{Timelike}, we do an analogous investigation of the timelike case. Our ansatz for the 
auxiliary fields gives rise to three choices for the vector field components and one of them leads to a new set of solutions that exists only in GMG and not in NMG or TMG. Additionally,  
we also find some solutions that were overlooked in earlier works \cite{Alkac:2015lma, Deger:2013yla}. We conclude in section \ref{Conc} with some 
comments and future directions. In appendix \ref{app1} we verify that the new AdS vacuum with non-zero vector fields is 1/4 supersymmetric whereas the other 
one with vanishing vector fields preserve full supersymmetry. In appendix \ref{app2} we give a preliminary analysis of the black hole like geometry that one obtains from our timelike squashed AdS solution (\ref{hopf2}).

\begin{table}
\begin{center}
\begin{tabular}{lcccc}
\hline
Solution & Conditions & GMG & NMG & TMG
\\
\hline
Round AdS \, (max. susy)
& $V_1=V_0=0, A\neq 0$ 
& (\ref{RoundAdS}) 
& \checkmark 
& \checkmark
\\
Round AdS \, (1/4 susy)
& $V_1 = V_0 =-A \neq 0$ 
& (\ref{hopf})
&\checkmark
&\xmark
\\
Spacelike squashed AdS 
& $V_1=A\neq 0, V_0 \neq 0, V^2>0$ 
& (\ref{warpedads}) 
& \checkmark
& \checkmark
\\
Timelike stretched AdS 
& $V_1=A\neq 0, V_0 \neq 0, V^2<0$ 
& (\ref{warpedads}) 
& \checkmark
& \checkmark
\\
Null warped AdS
& $ V_1=A\neq 0, V^2=0$ 
& (\ref{warpedads}) 
& \checkmark
& \checkmark
\\
$AdS_2 \times \mathbb{R}$
& $V_1=A \neq 0, V_0=0$
& (\ref{AdS2})
& \checkmark
& \xmark
\\
Timelike warped flat
& $V_1=A=0 \, , \, V_0 \neq 0 $ 
& (\ref{warpedflat})
&\checkmark
&\checkmark
\\
AdS pp-wave
& $|V_1|=|V_0| \neq |A|$, $A \neq 0$
& (\ref{adspp})
& \checkmark
& \checkmark
\\
Flat space pp-wave
& $|V_1|=|V_0| \neq 0$, $A=0$
& (\ref{flatpp})
& \checkmark
& \checkmark
\\
Stationary Lifshitz 
& $V_0=\frac{m^2}{3\mu}, V_1 \neq -A, V_1 \neq 0$  
& (\ref{lif})
& Static
& \xmark
\\
Timelike warped AdS 
& $V_0=\frac{m^2}{3\mu}, V_1=-A\neq 0, V^2 \neq 0$  
& (\ref{hopf2})
& $\mathbb{R}_t \times H_2  $
& \xmark
\\
A deformation of AdS 
& $V_0=\frac{m^2}{3\mu}, V_1=0, A\neq 0$  
& (\ref{deform})
& \xmark
& \xmark
\\
\hline
\end{tabular}
\end{center}
\caption{Supersymmetric timelike solutions of  $\cN = (1,1)$ GMG in comparison with $\cN = (1,1)$ NMG and $\cN = (1,1)$ TMG. For GMG we assume 
$\sigma \neq 0, 1/\mu\neq0$ and $1/m^2 \neq 0$. When $A=0$, then $M=0$. }
\label{table}
\end{table}

\section{\texorpdfstring{$\cN = (1,1)$\, G} General Massive Supergravity}{\label{GMG1}}

We begin with a summary of the  $\cN = (1,1)$ General Massive Supergravity which was constructed in \cite{Alkac:2014hwa}.
The bosonic part of its Lagrangian is given by
\bea
e^{-1} {\mathcal L}_{\rm GMG} &=&  \s (R+2V^2 -2|S|^2 ) + 4 M A
\nn\\[.1truecm]
&-&\frac{\varepsilon^{\mu\nu\rho}}{4\mu}\,\left[ R_{\mu\nu}{}^{ab} \omega_{\rho ab}
+ \ft23 \omega_\mu^{ab} \omega_{\nu b}{}^c\omega_{\rho ca}
-4 F_{\mu\nu}V_\rho \right]
\nn\\[.1truecm]
&+& \frac{1}{m^2} \Big[ R_{\mu\nu} R^{\mu\nu} -\ft38 R^2   - R_{\mu\nu} V^\mu V^\nu -  F_{\mu\nu} F^{\mu\nu} +\ft14 R(V^2 -B^2)
\nn\\[.1truecm]
&& \quad\quad + \ft16 |S|^2(A^2-4B^2) -\ft12 V^2 (3A^2+4B^2) - 2V^\mu  B\partial_\mu A \Big]\,,
\label{GMG}
\eea
where $(\s,M, m^2,\mu)$  are arbitrary real constants and the complex scalar field $S$ is defined as $S=A+\rmi B$. 
The limit $m^2 \rightarrow \infty$ corresponds to $\cN=(1,1)$ TMG and the limit $\mu \rightarrow \infty$ corresponds to $\cN=(1,1)$ NMG models.
Their supersymmetric solutions were studied in \cite{Deger:2013yla} and \cite{Alkac:2015lma}, respectively. The model can be truncated to 
$\cN=1$ GMG \cite{Andringa:2009yc, Bergshoeff:2010mf} by setting the vector field $V$ and the imaginary part of the scalar field, i.e. $B$, to zero.  

Equations of motion for $A, B, V_\m$ and $g_{\m\n}$ fields are given respectively as,
\begin{eqnarray}
0 &=& 4 M - 4\s A + \frac{1}{m^2} \left[ \frac{2}{3} A^3 - B^2 A - 3V^2 A + 2 \left( \nabla \cdot V \right) B + 2 V^{\mu} \partial_{\mu} B \right]  \,,  \nn \\
0 &=& 4\s B + \frac{1}{m^2} \left[ \frac{1}{2} R B + A^2 B + \frac{8}{3} B^3 + 4 V^2 B + 2 V^{\mu} \partial_{\mu} A \right] \,,  \nn \\
0 &=&  4\s V_{\mu} - \frac{1}{m^2} \left[ 2 R_{\mu\nu} V^{\nu} + 4 \nabla^{\nu} F_{\mu\nu} + V_{\mu} \left( 3A^2 + 4B^2 - 
\frac{R}{2} \right) + 2 B\partial_{\mu} A \right] +\frac{2}{\mu} \varepsilon_{\mu\nu\rho}F^{\nu\rho} \, , \nn \\
0 &=&  \s \Big( R_{\mu\nu} +2  V_{\mu}V_{\nu} - \frac{1}{2} g_{\mu\nu} [ R + 2V^2 - 2(A^2 + B^2) ] \Big) -2 g_{\mu\nu} M A   + \frac{1}{\mu}C_{\mu\nu}  \label{fieldequations}
\\
&+&   \frac{1}{m^2} \Bigg[ \Box R_{\mu\nu} - \frac{1}{4} \nabla_{\mu}\nabla_{\nu} R + \frac{9}{4} R R_{\mu\nu} - 4 R^{\rho}_{\mu} R_{\nu \rho} - 2 F_{\mu}{}^{\rho} F_{\nu \rho}  \qquad \nn \\
&& \qquad + \frac{1}{4} R V_{\mu}V_{\nu} - 2 R_{\rho (\mu} V_{\nu)} V^{\rho} - \frac{1}{2} \Box (V_{\mu}V_{\nu}) + \nabla_{\rho} \nabla_{(\mu} (V_{\nu)} V^{\rho}) \qquad \nn \\
&& \qquad + \frac{1}{4} R_{\mu\nu} (V^2 - B^2) - \frac{1}{4} \nabla_{\mu} \nabla_{\nu} (V^2 - B^2) - \frac{1}{2} V_{\mu}V_{\nu} (3A^2 + 4B^2) \qquad \nn \\
&&  \qquad - 2B V_{(\mu} \partial_{\nu)} A - \frac{1}{2} g_{\mu\nu} \Big( \frac{13}{8} R^2 + \frac{1}{2} \Box R - 3 R_{\rho \sigma}^2 - R_{\rho \sigma} V^{\rho} V^{\sigma}  
\qquad \nn \\
&&  \qquad + \nabla_{\rho} \nabla_{\sigma} (V^{\rho} V^{\sigma}) - F_{\rho \sigma}^2 + \frac{1}{4} R (V^2 - B^2) - \frac{1}{2} \Box (V^2 - B^2) \qquad \nn \\
&&  \qquad + \frac{1}{6} (A^2 + B^2) (A^2 - 4B^2) - \frac{1}{2} V^2 (3A^2 + 4B^2) - 2 B V^{\rho}\partial_{\rho} A \Big) \Bigg] \nn 
\,,
\end{eqnarray}
where the Cotton tensor is defined as
\be
C^{\mu}_{\,\,\, \nu}=\varepsilon^{\mu\rho\sigma}\nabla_{\rho}(R_{\sigma\nu} - \frac{1}{4}g_{\sigma\nu}R) \, .
\ee
Note that the gravitational Chern-Simons term has no contribution to the scalar field equations.

The $\cN=(1,1)$ supersymmetry transformations are \footnote{In this paper, we follow the conventions of \cite{Alkac:2015lma}. In 
\cite{Deger:2013yla}, on the other hand, $S=-Z$ and $\sigma = 1$.}
\bea
\delta e_\mu{}^{a}  &= & \ft{1}{2}\bar{\epsilon}\gamma^{a}\psi_{\mu}+h.c.\,,
\nn\\[.1truecm]
\delta\psi_\mu  &= & (\partial_{\mu}+\ft14\omega_{\mu}{}^{ab}\,\gamma_{ab})\epsilon-\ft{\rmi}{2}  V_{\nu}\,\gamma^{\nu}\gamma_{\mu}\,\epsilon
-\ft S2 \gamma_\mu \epsilon^*\,,
\nn\\[.1truecm]
\delta V_\mu  &= & \ft{\rmi}{8}  \bar{\epsilon}\,\gamma^{\nu\rho}\gamma_{\mu}\left(2D_{[\nu}\psi_{\rho]}-{\rmi} 
V_{\sigma}\gamma^{\sigma}\gamma_{\nu}\,\psi_{\rho}-S\gamma_\n \psi_{\rho}^* \right)+h.c.\,,
\nn\\[.1truecm]
\delta S  & = & -\ft14 \bar{\epsilon^*}\,\gamma^{\mu\nu}\left(2D_{[\mu}\psi_{\nu]}
-\rmi V_{\s}\,\gamma^{\s}\gamma_{\mu}\psi_{\nu}-S\gamma_{\mu} \psi_{\nu}^*\right)\, ,
\label{TransformationRules}
\eea
where $\epsilon$ is a complex Dirac spinor. These  are off-shell transformations since the supersymmetry algebra closes without imposing the field equations
(\ref{fieldequations}).

The model has a fully supersymmetric $AdS_3$ vacuum when
\be
A= -\frac{1}{\ell}\, , \, B=0 \, , \, V_{\mu}=0 \, , \, M = A\sigma -\frac{A^3}{6m^2} \, ,
\label{fullads}
\ee
where the effective cosmological constant is $\Lambda =-1/\ell^2$, that is $R_{\mu\nu}= 2 \Lambda g_{\mu\nu}$.
Linearizing the theory around this vacuum, one finds that generically, graviton has 2 massless modes with $\eta=1$ and $\eta=-1$ and
two massive modes with masses $\eta_1$ and $\eta_2$ given by:
\be
\eta_1 \eta_2 =\frac{1}{\Omega} \, , \quad \eta_1 + \eta_2 = \frac{\ell m^2}{\mu \Omega} \, ,
\ee
where $\Omega = \sigma \ell^2 m^2 - \frac{1}{2}$. When mass values are repeated logarithmic modes appear and such points of the parameter space are 
labeled as {\it critical}. Assuming that $1/\mu \neq 0$, there are 5 possibilities \cite{Bergshoeff:2010iy}:
\bea\nn
&&\textrm{i)} \, \eta_1=\eta_2 \neq \pm 1 \, , \quad \textrm{ii)} \, \eta_1= 1, \eta_2 \neq \pm 1 \, , \quad
\textrm{iii)} \, \eta_1= -1,  \eta_2 \neq \pm 1 \, , \\ 
&&\textrm{iv)} \, \eta_1=\eta_2 = 1 \, ,  \quad \textrm{v)} \, \eta_1=\eta_2 = -1 \, , 
\label{critical}
\eea
Supermultiplet structure of this theory at these critical points as well as at ordinary points were studied in \cite{Deger:2018pzj}.
 
Now we would like to find supersymmetric solutions of this model. Since it is off-shell, the Killing spinor analysis done 
in \cite{Deger:2013yla} is also valid here, which we summarize in the next two sections. As usual,
assuming the existence of at least one Killing spinor, one finds that there is a Killing vector constructed as a spinor bilinear which is either 
null or timelike.

\section{The Null Killing Vector}{\label{Null}}
The Killing spinor analysis of \cite{Deger:2013yla} shows that in the null case the vector field should be of the form $V_{\mu} = \partial_{\mu} \theta$
for some arbitrary function $\theta(u,x)$. Hence, the contribution of the gravitational Chern-Simons term to the vector field equation (\ref{fieldequations}) 
vanishes automatically. Furthermore, if $S$ is a real constant one finds that  
supersymmetry actually requires the vector field
to vanish and the Killing spinor becomes a constant spinor \cite{Deger:2013yla, Alkac:2015lma}. Now, we choose
\be
A=-1 \, , \, B=0 \, , \, V_{\mu}=0 \, ,
\ee
where the $AdS$ radius $|\ell|$ is fixed to 1 in (\ref{fullads}), which requires $M=\frac{1}{6m^2}-\sigma$ , from the scalar field equation. 
The only remaining field, namely the metric has the form \cite{Deger:2013yla}
\bea
ds^2 = dx^2 + 2\, e^{2x}\, du\, dv + Q(x,u)\, du^2 \, ,
\label{ppwavemetric}
\eea
with the null Killing vector in the $v$-direction. This generically describes a pp-wave, however when $Q(x,u)=\textrm{const.}$ \, or \, $Q(x,u)=e^{2x}$,   it 
is $AdS_3$. The metric field equation in (\ref{fieldequations}) implies that
\bea
Q_{xxxx} - \left(4 + \frac{m^2}{\mu}\right)Q_{xxx} + \left(\frac{9}{2} + m^2\sigma + \frac{3m^2}{\mu}\right)Q_{xx}
- \left(1 + 2m^2\sigma + \frac{2m^2}{\mu}\right)Q_x=0 \, .
\label{DiffEqQ}
\eea
When there is no degeneracy, the most general solution of this differential equation
is 
\bea\label{Qsolution}
Q(x,u) = c_1(u) + c_2(u)e^{2x} + c_3(u)e^{\lambda_1x} + c_4(u)e^{\lambda_2x} \,,
\eea
where the functions $c_i (u)\,,i=1,\cdots,4,$ are arbitrary functions of $u$ and 
\be
\lambda_{1,2} = 1 + \frac{m^2}{2\mu} \pm \sqrt{\left(1+\frac{m^2}{2\mu}\right)^2 -\left(\frac{1}{2} + \frac{m^2}{\mu} +m^2\sigma\right)} \, .
\ee
One can show that functions $c_1(u)$ and $c_2(u)$ can be set to zero without loss of generality \cite{Gibbons:2008vi, Alkac:2015lma}.

There are 5 special cases that must be analyzed separately \footnote{We ignore the case 
$\l_1=2$ and $\l_2=0$ since that requires $1/\mu=0$, which we don't allow.}. They correspond precisely to the critical 
points that we listed in (\ref{critical}) since $\lambda_i = \Omega \eta_i +1 , \, (i=1,2)$. Now solutions at these critical points take the form:
\bea
&&\textrm{i)} \l_1=\l_2=1 + \frac{m^2}{2\mu} ,  \mu^2=\frac{m^4}{4m^2\s -2}  , Q(x,u) = c_1(u) + c_2(u)e^{2x} + [c_3(u) + c_4(u)x]e^{\lambda_1x} , \nn \\
&&\textrm{ii)} \l_1=2  , \,  \l_2= \frac{m^2}{\mu} = \frac{1}{2} +m^2\s  , \, Q(x,u) = c_1(u) + [c_2(u) + c_3(u)x]e^{2x} + c_4(u)e^{\lambda_2x} \, , \nn \\
&&\textrm{iii)} \l_1=0 ,   \l_2= 2 + \frac{m^2}{\mu} = \frac{3}{2} - m^2\s , \, Q(x,u) = c_1(u) + c_2(u)e^{2x} + c_3(u)x + c_4(u)e^{\lambda_2x}  , \nn \\
&&\textrm{iv)} \l_1=\l_2=2  , \, m^2=\frac{3}{2\s}=2\mu  , \, Q(x,u) = c_1(u) + [c_2(u) + c_3(u)x + c_4(u)x^2]e^{2x}   \, , \nn \\
&&\textrm{v)} \l_1=\l_2=0  ,  \, m^2=\frac{3}{2\s}=-2\mu  , \, Q(x,u) = c_1(u) + c_2(u)e^{2x} + c_3(u)x + c_4(u) x^2  \, .
\eea
Note that in the last two cases we have triple degeneracy which does not occur in supersymmetric null solutions of $\cN=(1,1)$ NMG \cite{Alkac:2015lma}. 

In general, these solutions preserve 1/4 supersymmetry except the round $AdS_3$ which preserves full supersymmetry \cite{Alkac:2015lma}.

\section{The Timelike Killing Vector}{\label{Timelike}}

We now move on to the timelike case. We will employ the same ansatz for scalar and vector fields as in \cite{Deger:2013yla} and \cite{Alkac:2015lma}, namely
\begin{equation}
A = \mbox{const.} \, , \, B=0 \, , \, V_0 = \mbox{const.} \, , \, V_1 = \mbox{const.} \, , \, V_2 = 0 \, ,
\label{config}
\end{equation}
where tangent indices \{0,1,2\} correspond to $\{t,x,y\}$ coordinates, respectively. The metric with a timelike Killing vector in the $t$-direction 
has the form 
\cite{Deger:2013yla}
\be
ds^2= - e^{2\varphi(y)} \left(dt+ C(y)\, dx \right)^2 + e^{2\l(y)} (dx^2+dy^2)\ ,
\label{metric}
\ee
where $\lambda(y)$, $\vf(y)$ and  $ C(y)$ are arbitrary functions of $y$. For supersymmetry, metric functions should satisfy
\begin{eqnarray}
e^{-\l}\varphi' & = & V_{1}+A \, , \nn \\[.1truecm]
\label{Constraints}
e^{-\l}\l' & = & A-V_{1} \, ,  \\[.1truecm]
e^{-\l} C' & = & 2V_{0}\,e^{\l-\varphi} \nn \, ,
\end{eqnarray}
where prime indicates differentiation with respect to $y$. Note that, for this differential equation system the following choices are special 
\be
\textrm{i}) \, V_1=A \, , \quad \textrm{ii})\, V_1=-A \, , \quad \textrm{iii})\, V_0=0 \, , \quad \textrm{iv})\, V_1=0 \, , 
\label{special}
\ee
since they reduce the number of independent metric functions from three to two or one.
The significance of \, $V_1=-A$ case was overlooked in \cite{Alkac:2015lma}, and hence the corresponding solutions were missed. Also, cases with $A=M=0$ were
not considered in \cite{Alkac:2015lma} and \cite{Deger:2013yla} since this makes the effective cosmological constant zero. We will allow this option too.

Next, we have to solve the field equations of the model (\ref{fieldequations}) with these constraints. The advantage of our ansatz (\ref{config})
is that it makes scalar and vector field equations algebraic, and once they are satisfied Einstein equations become 
automatic. From the two non-trivial vector field equations one can show that the condition
\be
\left(\frac{3V_0}{m^2}-\frac{1}{\mu}\right)(V_1-A)(V_0^2-V_1^2)=0 \, ,
\ee
has to be fulfilled, which puts a strong restriction on the possible supersymmetric solutions. After satisfying this condition, 
only one of the vector field equations in (\ref{fieldequations}) remains free. Now we will follow
these three possibilities and solve the constraint equations (\ref{Constraints}) together with the remaining vector field equation for each case, paying
particular attention to the special choices (\ref{special}).

Supersymmetric solutions in this section are $1/4$ supersymmetric \cite{Deger:2013yla}, except for the round $AdS_3$ with no vector fields (\ref{RoundAdS}) which
is fully supersymmetric as we show in appendix \ref{app1}.

\subsection{The \texorpdfstring{$V_1 = A$\, c} case}

In this case the scalar field equation implies
\be
M=A\s + \frac{A}{4m^2}\left(\frac{7A^2}{3}-3V_0^2\right) \, ,
\label{sc1}
\ee
and the remaining vector field equation becomes
\be
\sigma=\frac{2V_0}{\mu} + \frac{7}{4m^2}\left(A^2-3V_0^2\right) \, .
\label{vec1}
\ee
There are two special cases (\ref{special}), namely $V_0=0$ and $V_1=A=0$.

\subsubsection{The \texorpdfstring{$V_1=A \neq 0$ \, c} case} \label{v1a}

In this case solving equations (\ref{Constraints}) the metric (\ref{metric}) takes the form
\be
ds^2=\frac{V^2}{A^2}\left(dx+ \frac{V_0A}{V^2}e^{2Ay}dt\right)^2 - \frac{A^2}{V^2}e^{4Ay} dt^2 + dy^2 \, ,
\label{warpedads}
\ee 
where $V^2 \equiv -V_0^2+V_1^2$. 

When $V_0 \neq 0$ this corresponds to a warped $AdS_3$ space with warping parameter
\be
\nu^2=1- \frac{V^2}{A^2} \, .
\ee
Depending on the norm of the vector field, i.e. $V^2$, being positive, negative or zero, it describes a
spacelike squashed ($0 <\nu^2 <1$), a timelike stretched ($\nu^2>1$) or a null warped $AdS_3$ spacetime, respectively (see \cite{Deger:2013yla, Alkac:2015lma} for details).

When $V_0=0$, note that (\ref{vec1}) requires $\s\geq 0$ and $M=7A^3/(3m^2)$. The metric (\ref{warpedads}) takes the form
\be
ds^2=-e^{4Ay}dt^2 +dy^2+dx^2 \, ,
\label{AdS2}
\ee
which corresponds to $AdS_2 \times \mathbb{R}$ or $AdS_2 \times S^1$ geometries. Note that the gravitational 
Chern-Simons term has no effect. Here, if we assume $\sigma \neq 0$ then this solution does not exist in $\cN=(1,1)$ TMG.

\subsubsection{The \texorpdfstring{$V_1=A=0$ and $V_0 \neq 0$ \, c} case} \label{v1a0}

For this case $M=0$ and (\ref{vec1}) implies  
$V_0= \frac{2m^2}{21}\left(\frac{2}{\mu} \pm \sqrt{\frac{4}{\mu^2}-\frac{21\s}{m^2}}\right)$. 
Solving supersymmetry constraints (\ref{Constraints}) we get
\be 
ds^2=-(dt+2V_0ydx)^2 +dx^2 +dy^2 \, ,
\label{warpedflat}
\ee
which represents a timelike warped flat space \cite{Deger:2016vrn}. Note that $V_0=0$ would imply $\sigma=0$ from (\ref{vec1}) which we don't allow.

This is still a solution when $1/\mu=0$ or $1/m^2=0$, but was not considered in \cite{Alkac:2015lma} and \cite{Deger:2013yla}.

\subsection{The \texorpdfstring{$|V_1| = |V_0| \, , \,  V_1 \neq A $\,  c} case}

Let  $V_0 = -\varepsilon V_1$ where $\varepsilon^2=1$.
The scalar and remaining vector field equation give
\bea
\label{sca2}
M &=& A\s - \frac{A^3}{6m^2} \, , \\
0 &=& V_1\left[\frac{(A^2 + 4AV_1 + 2V_1^2)}{m^2} - \frac{2\varepsilon}{\mu}(A+V_1) + 2\s \right] \, .
\label{vec2}
\eea
Here the special cases (\ref{special}) to be considered are $V_1=0$ and $V_1=-A$.

\subsubsection{The \texorpdfstring{$V_1\neq -A $\, c} case}

For this case solving (\ref{Constraints}) gives the metric 
\be
ds^2=-y^{2\a}dt^2 + \frac{2\varepsilon}{A-V_1}y^{\a -1}dtdx + \frac{dy^2}{y^2(A-V_1)^2} \, ,
\label{adspp}
\ee
where $\a =(V_1+A)/(V_1-A)$. Although, at first sight $V_1=A$ looks problematic, in this subsection that is not allowed. 
For $A \neq 0$, this solution corresponds 
to an $AdS_3$ pp-wave when $|V_1|=|V_0| \neq 0$ \cite{Deger:2013yla, Alkac:2015lma}. Supersymmetry requires 
$\varepsilon =-1$ \cite{Deger:2013yla}.

When $A=0$, from (\ref{sca2}) we have $M=0$, and 
\be
V_1=-\frac{m^2\varepsilon}{2\mu} \pm \sqrt{m^2\s - \frac{m^4}{4\mu^2}} \, ,
\ee
which requires $m^2 \leq 4\mu^2\s$. The metric (\ref{adspp}) with $\alpha=1$ and $V_1 \neq 0$ becomes
\be
ds^2= -y^2dt^2 - \frac{2\varepsilon}{V_1}dtdx + \frac{dy^2}{V_1^2y^2} \, .
\ee
After the coordinate transformations $u=(\ln y)/V_1$ and $z=-\varepsilon x/V_1$, we get
\be
ds^2= - e^{2uV_1}dt^2 + 2dtdz + du^2 \, ,
\label{flatpp}
\ee
which describes a pp-wave in flat spacetime in Brinkmann coordinates \cite{Deger:2016vrn}.
This solution also exists in TMG and NMG but was not considered in \cite{Deger:2013yla} and \cite{Alkac:2015lma}.

On the other hand,
when $A \neq 0$ but $|V_1|=|V_0|=0$, the metric (\ref{adspp}) takes the form
\begin{equation}
ds^{2}=\frac{1}{A^2y^{2}}\,(-d\tau^{2}+dx^{2}+dy^{2}) \, ,
\label{RoundAdS}
\end{equation}
which is the round $AdS_3$ spacetime with radius $1/|A|$. Here we defined $\tau=At -\varepsilon x$. This is the
only solution which is maximally supersymmetric as we show in appendix \ref{app1}.

\subsubsection{The \texorpdfstring{$V_1=-A $\, c} case}

Note that in this case (\ref{vec2}) and (\ref{sca2})  imply $M=2A\s/3$ and $A^2=2m^2\s$, which means $\s>0$.
Now, putting $\alpha=0$ in (\ref{adspp}) we obtain
\be
ds^2=-\left(  dt-\frac{\varepsilon dx}{2Ay}\right)^2 +\frac{1}{4A^2y^2}(dx^2 + dy^2) \, ,
\label{hopf}
\ee
which is the round $AdS_3$ with radius $1/|A|$ written as a timelike Hopf fibration over a 
hyperbolic space without any warping \cite{Deger:2016vrn}. Since AdS is conformally flat, its
Cotton tensor vanishes. Therefore, this solution also exists in NMG which
was not noticed in \cite{Alkac:2015lma}. It is not a solution of TMG.

In appendix \ref{app1} we show that unlike our previous $AdS_3$ solution (\ref{RoundAdS}) this one preserves only 1/4 supersymmetry.
This requires $\varepsilon =-1$, that is $V_0=V_1$, and the resulting Killing spinor is constant.

\subsection{The \texorpdfstring{$V_0 = \frac{m^2}{3\mu} \, , \, |V_0|  \neq |V_1| \, , V_1 \neq A$ \,  c} case}

The main difference between the solutions of $\cN = (1,1)$ GMG and NMG  \cite{Alkac:2015lma} shows up in this class
since $1/\mu$ appears directly in $V_0$. Solutions here either don't exist in NMG or they exist as a special case of the more 
general form that is allowed in GMG. They are not solutions of $\cN = (1,1)$ TMG \cite{Deger:2013yla}.

The scalar and vector field equations become
\bea
M&=&A \left(\s + \frac{3V_1^2}{4m^2} -\frac{m^2}{12\mu^2}-\frac{A^2}{6m^2}\right) \, , \\
0&=& m^4 +3\mu^2(5V_1^2+4AV_1-2A^2) -12m^2\mu^2\s \, .
\eea
The cases $V_1=-A$ and $V_1=0$ should be considered separately (\ref{special}).

\subsubsection{The \texorpdfstring{$V_1 \neq 0 \, $\,  c} case}

For this case solving (\ref{Constraints}) leads to the metric
\be
ds^2=-y^{2\a}dt^2 - \frac{V_0(\alpha+1) y^{\a -1}}{V_1^2}dtdx + \frac{V^2(\alpha+1)^2}{4V_1^4}\frac{dx^2}{y^2} +
\frac{(\alpha +1)^2}{4V_1^2}\frac{dy^2}{y^2} \, ,
\label{lif}
\ee
where $\a =(V_1+A)/(V_1-A)$. Note that, since  $V_1 \neq 0$ we have $ \a \neq -1$.

First let us assume that $V_1 \neq -A$, which implies that $\a \neq 0$. Then, the metric (\ref{lif}) remains invariant
under the rescalings $y \rightarrow \l y, x \rightarrow \l x, t \rightarrow
\l^{-\a} t$ where $\l$ is an arbitrary constant. Hence this solution corresponds to a stationary Lifshitz spacetime \cite{Sarioglu:2011vz}
with dynamical exponent $z=-\a$. Such a solution was obtained before for Minimal Massive 3D Gravity model \cite{Bergshoeff:2014pca} in \cite{Charyyev:2017uuu}. The solution exists even when $A=0$ with dynamical exponent $z=-1$, although $M=0$. In the NMG limit, namely $\mu \rightarrow \infty$, we have $V_0=0$ and (\ref{lif}) becomes static Lifshitz spacetime
as in minimal massive 3D gravity \cite{Charyyev:2017uuu}.

When $V_1= -A$, i.e. $\alpha=0$ and $M=A^3/(3m^2)$, the metric (\ref{lif}) with $\tau=2A^2t/V_0$ becomes
\be
ds^2=\frac{1}{4A^2}\left[-\nu^2\left( d\tau+\frac{dx}{y}\right)^2 +\frac{1}{y^2}(dx^2 + dy^2)\right] \, ,
\label{hopf2}
\ee
which is a timelike  warped $AdS_3$ \cite{Deger:2016vrn} with warping $\nu^2=V_0^2/A^2$. The form of the metric is 
similar to our solution given in (\ref{hopf}), however here we have $V_0 \neq \pm A$, therefore it is not round $AdS_3$. Moreover, unlike our previous timelike warped 
solution (\ref{warpedads}) there is no restriction 
on the warping, it can be squashed or stretched. When it is squashed, it becomes a self-dual type solution \cite{Coussaert:1994tu} after an appropriate identification
that is free from closed timelike curves and has a Killing horizon \cite{Anninos:2008fx} which we study further in appendix \ref{app2}. Note that in the NMG limit, that is
$1/\mu=0$, $V_0$ vanishes and the metric (\ref{hopf2}) becomes $\mathbb{R}_t \times H_2$, where $H_2$ is 
a 2-dimensional hyperbolic space. This case was overlooked in \cite{Alkac:2015lma}.

\subsubsection{The \texorpdfstring{$V_1= 0$\, c} case}
In this case $M=-2A^3/(3m^2)$ and $A^2=m^2(m^2-12\mu^2\s)/6\mu^2$, therefore we need to have $m^2 > 12 \mu^2\s$.
Here we also assume $A\neq 0$ (that is, $m^2 \neq 12 \mu^2 \s$) since this was covered
in subsection \ref{v1a0}.
Solving (\ref{Constraints}) the metric becomes
\be
ds^2= -\frac{1}{y^2}\left[dt + \frac{2m^2}{3\mu A^2} \ln y dx\right]^2 + \frac{1}{A^2y^2}(dx^2+dy^2) \, .
\label{deform}
\ee
This can be thought of some `logarithmic' deformation of $AdS_3$ in Poincar\'e coordinates. We are not familiar with this metric.
Two of its curvature invariants are
\be
R=2(V_0^2 - 3A^2) \, , \quad R_{\mu\nu}R^{\mu\nu}=4(3V_0^4-4A^2V_0^2+3A^4) \, .
\ee

\section{Discussion}\label{Conc}

In this paper we constructed a large number of supersymmetric backgrounds of $\cN=(1,1)$ GMG theory \cite{Alkac:2014hwa}. 
Since supersymmetric solutions of  $\cN=(1,1)$ TMG and NMG were studied earlier \cite{Deger:2013yla, Alkac:2015lma} with the same ansatz for auxiliary fields,
the picture is now complete and one can see consequences of including separate off-shell invariant pieces from our table \ref{table}.
In \cite{Alkac:2015lma}
it was found that $\cN=(1,1)$ NMG allows more solutions in comparison to $\cN=(1,1)$ TMG \cite{Deger:2013yla} and here we showed that 
$\cN=(1,1)$ GMG is even richer. Since the difference between NMG and GMG is the presence of the gravitational Chern-Simons term in the latter,
our findings highlight the effect of this term. In particular, we have seen that the static Lifshitz solution of NMG becomes stationary in GMG. This phenomena
was also observed for Minimal Massive Gravity \cite{Bergshoeff:2014pca} in \cite{Charyyev:2017uuu}. Similarly, comparing solutions of GMG with TMG in table \ref{table},
one can see the significance of inclusion of higher derivative terms. 

Looking at supersymmetric solutions \cite{Deger:2016vrn} of a closely related model, 
namely $\cN=(2,0)$ TMG \cite{Alkac:2014hwa} one realizes that many are in common and almost all of them are homogeneous backgrounds \cite{Charyyev:2017uuu}.
It is desirable to understand the connection between supersymmetry and homogeneity better.
As far as we know, (\ref{lif}) is the first example of a supersymmetric stationary Lifshitz spacetime \cite{Sarioglu:2011vz} and its properties should be investigated 
further. In particular, one may look for some stationary, supersymmetric Lifshitz black holes especially since, it is known that they don't exist
in $\cN=(1,1)$ NMG \cite{Alkac:2015lma}. It would also be interesting to study thermodynamics
and conserved charges \cite{Donnay:2015joa} of our timelike warped AdS solution (\ref{hopf2}) which has a Killing horizon as shown in appendix \ref{app2}.
We were able to give geometric identification of all our solutions except (\ref{deform}). It is a deformation of $AdS_3$ that we have not met before and it deserves 
a more detailed examination.

The second AdS vacuum that we found (\ref{hopf}) is worth investigating further. For example, one may try to obtain 
a renormalization group flow between the two AdS vacua using an appropriate string solution. For that purpose, constructing matter couplings or extensions with more number of supersymmetries of this theory might 
be necessary. Moreover, by making the $x$-coordinate periodic in (\ref{hopf}) one obtains a massless, static BTZ
black hole \cite{Banados:1992wn} with a non-zero charge. To check whether the model admits massive, rotating versions of this 
needs more elaborate investigation which will require relaxing our ansatz (\ref{config}). Searching supersymmetric flows between warped vacua of the model is also 
interesting \cite{Colgain:2015ela}. 

Finally, connecting these 3-dimensional models to higher dimensions as well as to 3-dimensional on-shell supergravities is an important task to do. 
We hope to come back to these issues in near future.

\section*{Acknowledgments}
ZN is fully, NSD and {\"O}S are partially supported by the Scientific and Technological Research Council of Turkey (T{\"U}B\.{I}TAK) Grant No.116F137. We thank 
D. Anninos for an e-mail correspondence. NSD wishes to thank the Abdus Salam ICTP for hospitality where some part of this paper was written.

\appendix
\section{Supersymmetry of \texorpdfstring{$AdS_3$\, w} with Vector Fields \label{app1}}
In this appendix, we show that our $AdS_3$ solution with the metric (\ref{hopf}) and non-zero vector fields  
$V_1=-A$ and $V_0=-\varepsilon V_1$ preserves 1/4 supersymmetry when $\varepsilon =-1$.

The Killing spinor equation $\delta \psi_{\mu} = 0$ can be written from (\ref{TransformationRules}) with $S=A$ as:
\bea
d \e +\ft14\omega{}_{ab}\,\gamma^{ab}\epsilon  \,
-\ft A2 \gamma_a e^a \epsilon^* -\kappa \ft{\rmi}{2}  (V_{a}\,\gamma^{a})(\gamma_{b} e^b) \epsilon = 0 \, ,
\label{KillingSpinor}
\eea
where we inserted the constant $\kappa$ (which actually is 1) to be able to compare with the $AdS_3$ solution that has no vector fields (\ref{RoundAdS}) more easily.

We choose the orthonormal frame for the metric (\ref{hopf}) as
\be 
e^0 = dt - \frac{\varepsilon}{2Ay}dx \, , \, e^1 = \frac{\varepsilon}{2Ay}dx \, , \, e^2 = \frac{\varepsilon}{2Ay}dy \, ,
\ee
whose spin connections are
\be
\omega_{01} = \frac {dy}{2y} \, , \, \omega_{02} = -\frac{dx}{2y} \, , \, \omega_{12} = - \varepsilon A dt  -\frac{dx}{2y} \, .
\ee
We take the $\gamma$-matrices as
\be
\gamma_0=i\sigma_2 \, , \, \gamma_1 =\sigma_1 \, , \, \gamma_2=\sigma_3 \, ,
\ee
where $\sigma_i$'s are Pauli matrices and decompose the complex Dirac spinor $\epsilon$ as
\be
\epsilon = \left( \begin{array}{cc} \epsilon_1 + i \zeta_1 \\ \epsilon_2 + i \zeta_2 \end{array} \right) \, .
\ee

With these choices, for $\varepsilon =-1$ we get the following 4 equations from (\ref{KillingSpinor}):
\bea
&& d\epsilon_1 - A\epsilon_2 dt =0 \, , \nn \\
&& d\zeta_1 - \frac{\zeta_1}{2y}dy =0 \, , \nn \\
&& d\epsilon_2 + A \epsilon_1 dt + \kappa \left(\frac{\zeta_1}{2y}dy - A\zeta_2dt\right) = 0 \, , \nn \\
&& d\zeta_2 + \frac{\zeta_2}{2y}dy -\frac{\zeta_1}{y}dx -\kappa \left(\frac{\epsilon_1}{2y}dy - A\epsilon_2dt\right)=0 \, .
\label{spinoreqns}
\eea
It is easy to see that for $\kappa = 0$ (that is pure $AdS_3$) this differential equation system has the following 4 linearly independent 
solutions:
\bea
&& \textrm{i}) \, \zeta_1 = \zeta_2=0 \, ,\, \epsilon_1=\cos At \, ,\, \epsilon_2=-\sin At \, , \nn  \\
&& \textrm{ii}) \, \zeta_1 = \zeta_2=0 \, ,\, \epsilon_1=\sin At \, , \, \epsilon_2=\cos At \, , \nn  \\
&& \textrm{iii}) \, \zeta_1=0 \, , \, \zeta_2 = y^{-1/2} \, , \, \epsilon_1=\epsilon_2=0 \nn  \, , \\
&& \textrm{iv}) \, \zeta_1=y^{1/2} \, , \, \zeta_2= x y^{-1/2} \, , \, \epsilon_1=\epsilon_2=0  \, . 
\label{spinorsoln}
\eea
So, $AdS_3$ with no vector fields is fully supersymmetric as it should be. Now,  for $\kappa = 1$ (so there is a contribution from the vector 
field) we see that Killing spinor equations (\ref{spinoreqns}) admit only one solution which is given as:
\be
\epsilon_0 = \left( \begin{array}{cc} 1 \\  i  \end{array} \right) \, \, , \, \,  (\varepsilon =-1)  \, .
\label{susysoln}
\ee
The constant Killing spinor (\ref{susysoln}) satisfies $\mathbb{P}  \epsilon_0 = \epsilon_0$ and $\mathbb{P}^*  \epsilon_0 =0$
where $\mathbb{P} = \frac{1}{2} (\mathbb{I}_2 + i\gamma^0)$.
 
On the other hand, when $\varepsilon =1$, the Killing spinor equations (\ref{spinoreqns}) still admit 4 solutions for
pure $AdS_3$  (just interchange $\zeta_i \leftrightarrow \epsilon_i$ in (\ref{spinorsoln})) as it should be, since 
$\varepsilon$ can be absorbed to the $x$-coordinate in the metric (\ref{hopf}). But there are no solutions with 
$\varepsilon=\kappa=1$. 

In summary, our $AdS_3$ solution with non-zero vector  fields (\ref{hopf}) preserves 1/4 supersymmetry when $\varepsilon =-1$. This 
result is in agreement with the Killing spinor analysis done in the appendix of \cite{Deger:2013yla}
where it was shown that supersymmetry enhancement happens only for $AdS_3$ with no vector fields turned on.

\section{Timelike Squashed Self-Dual Black Holes} \label{app2}

The spacelike (\ref{warpedads}) and timelike squashed solutions (\ref{hopf2}) that we obtained are examples of
so-called self-dual type solutions \cite{Coussaert:1994tu} after a proper identification. Although they do not have an 
event horizon or a singularity, they possess a Killing horizon and hence 
can be interpreted as black hole like objects with similar thermodynamic properties \cite{Anninos:2008fx}. The spacelike squashed self-dual solution 
appears as the near horizon region of the extremal Kerr black hole \cite{Bengtsson:2005zj} and has been studied for TMG 
in \cite{Chen:2010qm, Jugeau:2010nq, Li:2010zr}  and for $\cN=(1,1)$ extended TMG in \cite{Deger:2013yla}. However, so far the timelike
version has only appeared in $\cN=(2,0)$ TMG as a supersymmetric solution \cite{Deger:2016vrn}. Here, we initiate its study by looking
at its geometry  more closely.

The timelike $AdS_3$ solution that we obtained (\ref{hopf2}) is squashed when $V_0^2 < A^2$ which requires 
$m^2>12\mu^2 \s$. It is a self-dual type solution \cite{Coussaert:1994tu} with no closed timelike curves when the $x$-coordinate is 
identified such that $x\sim x+ 2\pi$. First, note that after the following coordinate changes:
\be
y=\frac{e^u}{\cosh \sigma} \, , \, x = e^u \tanh \sigma \, , \, \tau = t' - 2 \tan^{-1}[\tanh(\frac{\sigma}{2})] \, ,
\ee
it can be mapped into global $AdS$ coordinates
\be
ds^2 =\frac{1}{4A^2}\left[\cosh^2\sigma du^2 + d\sigma^2 - \nu^2(dt'+\sinh\sigma du)^2\right] \, ,
\label{global}
\ee 
where now the periodic coordinate is $u$. To see its causal structure more clearly, we now go to Schwarzschild type coordinates in (\ref{hopf2}) 
with $\nu^2<1$ by defining\footnote{Going to Schwarzschild type coordinates directly from (\ref{hopf2}) rather than 
(\ref{global}) leads to a much simpler set of transformations compared to those given in \cite{Anninos:2008fx} and \cite{Deger:2013yla}.}
\bea
y &=& \frac{1}{r-r_h} \, , \nn \\
\tau &=& \frac{\tilde{t}}{\nu} - \frac{6(1-\nu)r_h}{\nu(4-\nu^2)} \theta \, ,\nn \\
x &=&\frac{6}{4-\nu^2}\theta \, ,
\eea
where $r_h \geq 0$ is a constant. Now, the metric (\ref{hopf2}) transforms into
\be
ds^2=\frac{1}{4A^2}\left[\frac{[36(r-r_h)^2}{(4-\nu^2)^2} d\theta^2 + \frac{dr^2}{(r-r_h)^2} - 
 \left(d \tilde{t} + \frac{6(\nu r-r_h)}{4-\nu^2}d\theta\right)^2\right] \, ,
\ee
where $\theta \sim \theta + 2\pi$. Note that there is a Killing horizon at $r=r_h$ where the Killing vector $\chi = \frac{6(\nu r -r_h)}{4-\nu^2} \partial_{\tilde{t}} - 
\partial_{\theta}$ \, becomes null. Unlike the spacelike self-dual case \cite{Deger:2013yla}, here the Killing horizon has Lorentzian signature.

\end{document}